\documentclass[12pt]{article}
\usepackage{amsmath,amsfonts,cite}
\usepackage[all]{xy}
\usepackage{latexsym}
\usepackage{amssymb}
\usepackage{amscd}
\usepackage{amsmath}
\newtheorem{fact}{Fact}[section]
\newtheorem{theo}[fact]{Theorem}
\newtheorem{prop}[fact]{Proposition}
\newtheorem{lemma}[fact]{Lemma}

\newtheorem{defi}[fact]{Definition}

\newtheorem{step}{Step}

\newcommand{\qed}{$\hfill{\Box}$}
\newcommand{\mc}{\mathcal}
\newcommand{\re}{\mathbb{R}}

\newcommand{\ih}{\frac{i}{\hbar}}
\newcommand{\half}{\frac{1}{2}}

\newcommand{\moyall}{\ast_{\lambda}}
\newcommand{\starl}{\star_{\hbar}}

\begin{document}
\vspace{45 mm}
\title{Deformation Quantization in Singular Spaces}

\author{Cesar\ Maldonado-Mercado\\
\small Department of Mathematics,University Gardens, University of Glasgow\\
\small Glasgow, G12 8QW, United Kingdom}

\date{November, 2003}
\maketitle
\begin{abstract}
We present a method of quantizing analytic spaces $X$ immersed in
an arbitrary smooth ambient manifold $M$. Remarkably our approach
can be applied to singular spaces. We begin by quantizing the
cotangent bundle of the manifold $M$. Using a super-manifold
framework we modify the Fedosov construction in a way such that
the $\star$-product of the  functions lifted from the base
manifold turns out to be the usual commutative product of smooth
functions on $M$. This condition allows us to lift the ideals
associated to the analytic spaces on the base manifold to form
left (or right) ideals on $(\mc{O}_{\Omega^1 M}[[\hbar]],\starl)$
in a way independent of the choice of generators and leading to a
finite set of PDEs defining the functions in the quantum algebra
associated to $X$. Some examples are included.
\end{abstract}

\newpage

\sloppy

\hyphenation{super-manifolds}

\section{Introduction}

Deformation quantization is mathematically speaking a way of
defining non-commutative associative products on a Poisson
manifold, called $\star$-products, in a way such that the
non-commutativity is controlled by a deformation parameter. The
usual pointwise product is recovered as a limit case when this
deformation parameter is negligible and the Poisson structure is
found to be in the same fashion a limit case of the
$\star$-commutator in accordance with the quantum correspondence
principle.

Formally speaking, consider a smooth manifold $N$. A star-product
on $\mc{O}_N[[\hbar]]$ is an associative $\re[[\hbar]]$-linear
product
\begin{equation*}
f\starl g:= \sum_{k=0}^{\infty} (-\frac{i\hbar}{2})^{k}m_{k}(f,g)
\end{equation*}
where any $m_{k}$ is a bidifferential operator of finite total
order and $m_{0}(f,g)=fg$. It then can be shown that the operation
on $\mc{O}_{N}[[\hbar]]$ defined by $\{f,g\}=\lim_{\hbar
\rightarrow 0}\frac{1}{i\hbar}[f,g]$ is indeed a Poisson structure
on $M$. (For a review from the mathematical perspective see
\cite{Dito}.)

From the physics point of view deformation quantization is a new
autonomous reformulation of quantum mechanics. Although still in
development nowadays it is capable of reproducing numerous
examples from the ordinary operator formulation and has been found
to be closely related to the path integral formulation. (For a
review from the physics perspective see
\cite{Zachos}\cite{Hirshfeld}.)

One of the most powerful trends of work in mathematical physics
during the last century was the generalization of the formulation
of physical theories from the Euclidean case to the manifold
framework. In this way classical mechanics was  formulated in
terms of a Poisson structure generalizing the classical notion of
the Poisson bracket.  Another example of this has been the
generalization of the operator formulation of quantum mechanics to
the non-flat case. Different answers have been found e.g.
geometric quantization, group theoretic quantization. (For an
extended discussion see \cite{Isham}.)

Deformation quantization has followed a similar way going from the
first $\star$-product found by Moyal following physical ideas and
then generalized by Fedosov, who gave an explicit construction of
a $\star$-product in an arbitrary symplectic manifold
\cite{fedosovb}\cite{fedosov1}.

The quantization of singular spaces has been somewhat rejected
until very recently. This is one of the new possibilities provided
by deformation quantization, since the traditional approaches
breakdown in this case. Merkulov proposed in \cite{merkulov} a way
of quantizing algebraic varieties immersed in some $\mathbb{R}^n$.
This included the possibility of a non-empty set of singular
points.

In this work we provide a general construction to quantize
arbitrary analytic spaces (including the singular case), immersed
in any smooth analytic real manifold. (A different approach can be
found in \cite{landsman}). We summarize the general context of
this work  in the following table.

\begin{tabular}{|c|c|c|}
\hline
          & Euclidean & Manifold framework \\ \hline
  Classical Mechanics & Poisson bracket & Poisson structure \\
  \hline
  Quantum Mechanics. & Heisenberg's formulation & e.g. group quantization \\
  \hline
  Deformation Quantization & Moyal $\star$-product  & Fedosov construction
  \\ \hline
  Deformation Quantization & Merkulov's work &  This article \\

  of Singular spaces &  &   \\ \hline
\end{tabular}

\section{Quantization of Analytic Spaces}

Consider a smooth analytic manifold $M$ and $\mc{O}_M[[\hbar]]$,
the ring of formal power series with global analytic functions as
coefficients equipped with the usual commutative product.
Classically an analytic space $(X,\mc{O}_X[[\hbar]])$ immersed on
$M$ is defined by choosing a finitely generated vanishing ideal
$I$. Then the subspace $X$, which in general is not a smooth
submanifold, corresponds to the set of solutions of the system of
equations $\phi_{i}=0$, for any set $\phi_1,\ldots,\phi_n$ of
generators of $I$. The associated ring of functions is defined
then as $\mc{O}_X[[\hbar]]=\mc{O}_M[[\hbar]]/I$. (For a detailed
exposition see for example \cite{complex}).

Now consider the cotangent bundle $\pi:\Omega^1 M\rightarrow M$.
The ideal $I$ can be lifted via $\pi^{-1}(I)$ and in this way an
analytic space $(X,\mc{O}_X[[\hbar]])$ can be defined. This time
$\mc{O}_X[[\hbar]]=\mc{O}_{\Omega^1 M}[[\hbar]]/\pi^{-1}(I)$ and
$X\subset\Omega^1 M$. In physical terms this would describe a
system with a set of constrictions in the configuration space. Our
goal is to define a quantum version of this structure capable of
dealing with singular spaces.

Note that if one tries to replace naively the pointwise product of
functions by the star-product on $\Omega^1 M$ (which can be found
via the standard Fedosov construction for symplectic manifolds),
one finds two possible scenarios. If one fixes a set of generators
of the vanishing ideal $\pi^{-1}(I)$ before quantizing, the
construction will depend on this choice since the star product
depends on the smoothness around the vanishing points of the
possible choices of generators . On the other hand if one avoids
this choice and defines a left (or right) ideal and proceeds to
determine the normalizer, one finds an infinite number of
equations.

In our approach, we modify the Fedosov construction
\cite{fedosovb}\cite{fedosov1} in a way that alow us to find a
quantum algebra independently of the choice of generators and
corresponding to the solutions of a finite set of partial
differential equations.

The quantization procedure goes as follows

\begin{step}\textbf{Initial data.}
The necessary data to begin the construction is:
\begin{enumerate}
  \item {A smooth manifold $M$.}
  \item {A torsion free affine connection $\partial$ defined on
         $M$.}
  \item {The vanishing ideal $I$ associated to the classical analytic space.}
\end{enumerate}
\end{step}

\begin{step}\textbf{
Quantization of $\mathbf{\Omega^1 M.}$}
\end{step}
Before discussing  the details\footnote{For a related material on
the deformation quantization of cotangent bundles with a different
approach  see \cite{Bordemann}} we give in rough terms an overview
of this step. The goal is to construct a star product on the ring
of functions $\mc{O}_{\Omega^1 M}[[\hbar]]$ which coincides with
the pointwise multiplication when restricted to functions lifted
from the base, which is not the case in general for  the standard
Fedosov construction. The main tool is an auxiliary algebra
$\mc{W}$ where a star product $\starl$ can be defined in a
straightforward manner. Then a subalgebra
$\mc{W}_{D}\subset\mc{W}$ with a one to one correspondence that we
will denote as $\phi:\mc{O}_{\Omega^1 M}[[\hbar]]\rightarrow
\mc{W}_{D}$ is found. The star product $\starl'$ for
$\mc{O}_{\Omega^1 M}[[\hbar]]$ is defined as the one making the
following diagram commute
\[
\begin{CD}
\mc{O}_{\Omega^1 M}[[\hbar]]\otimes \mc{O}_{\Omega^1 M}[[\hbar]]
@>\starl'>> \mc{O}_{\Omega^1 M}[[\hbar]] \\ @ V \phi\otimes\phi VV @VV\phi V\\
\mc{W}_{D}\otimes \mc{W}_{D}  @>>\starl>\mc{W}_{D}
\end{CD}
\]
Then the key point in this step is to find the  subalgebra
$\mc{W}_{D}$, which turns out to be the set of flat sections of a
connection, and the correspondent map $\phi$.

Now we proceed with the  exposition in detail. We shall use  the
language of supermanifolds which makes it more simple and makes
the nature of the objects used more transparent.

Consider a ($3n|\,n$)-dimensional supermanifold
$\mc{M}:=\Omega^{1}M\times_{M}TM\times_{M}\Pi(TM).$ (Where we have
used the parity change operator $\Pi$.) For a coordinate system in
$\Omega^1M$ of the form $(x^{1}\ldots x^{n},p_{1}\ldots p_{n})$ we
have an associated coordinate system on $\mc{M}$ of the form
\begin{equation*}
(x^{1}\ldots x^{n},p_{1}\ldots p_{n},y^{1}\ldots
y^{n},\psi^{1}\ldots \psi^{n}).
\end{equation*}

\begin{defi} \label{weyl2}
 The Weyl algebra $\mc{W}$ on the supermanifold $\mc{M}$ is
the usual supercommutative algebra $\mc{O}_{\mc{M}}[[\hbar]]$,
 and  a typical element of $\mc{W}$ has locally the form
\begin{equation}
a(x,p,y,\psi)=\sum_{k,p,r=0}^{\infty} \hbar^{k} a_{k,i_{1}\ldots
i_{p},j_{1}\ldots j_{q}}^{k_{1}\ldots k_{r}}(x) y^{i_{1}}\ldots
y^{i_{p}} p_{k_{1}}\ldots p_{k_{r}} \psi^{j_{1}}\ldots
\psi^{j_{q}}, \label{locall}
\end{equation}
where the tensor $a_{k,i_{1}\ldots i_{p},j_{1}\ldots
j_{q}}^{k_{1}\ldots k_{r}}$ is symmetric in the $i_1\ldots i_p$
and $k_1\ldots k_r$ indices and antisymmetric in the $j_1\ldots
j_r$ indices.
\end{defi}

There is a natural $\star$-product defined on $\mc{W}$ given as
follows
\begin{equation*} f\starl
g=\exp (-\frac{i\hbar }{2}(\frac{\partial^{2}}{\partial
y^{a}\partial \tilde{p}_{ a}} -  \frac{\partial^{2}}{\partial
\tilde{y}^{a}\partial p_{ a}} ))f(x,p,y,\psi)g(x,\tilde{
p},\tilde{y},\psi)|_{p=\tilde{p},\,\,y=\tilde{y}}
\end{equation*}
where $f,g\in\mc{W}$. This product is manifestly covariant and is
easy to check it is associative. Note also that $f\starl g= g
\star_{-\hbar}f$.

\bigskip
Then some auxiliary vector fields on $\mc{M}$ are defined
\begin{eqnarray*}
i)&\delta:=&\psi^{a}\frac{\partial}{\partial y^{a}}\\
ii)&\delta^{\star}:= &y^{a}\frac{\partial}{\partial\psi^{a}}
\\ iii)& d:=&\psi^{a}\frac{\partial}{\partial
x^{a}}\\iv)&\delta^{-1}a:=&\frac{\delta^{*}}{p+q}a\\
v) &\partial a:=&\psi^{i}\partial_{i} a.
\end{eqnarray*}
where $a\in\mc{W}$ is an homogeneous element of order  $p$ in the
variable $y^{a}$ and of order $q$ in the anticommutative variable
$\psi^{a}$.
\begin{lemma}
Let $a\in\mc{W}$ be an homogeneous element as in the last
paragraph and define $a_{00}:=a(x,p,0,0)$, then
\begin{equation*}
\begin{array}{rcll}
i)& \delta^{2} a  & = & 0, \\
\rule{0pt}{3.5ex} ii) & \delta^{\star^2} a & = & 0,\\
\rule{0pt}{3.5ex} iii)& \delta a & = & - \ih[p_{j}\psi^{j},a]_* ,\\
\rule{0pt}{3.5ex} iv)& a & = & a_{00}+\frac{1}{p+q}(\delta
\delta^{\star} a+ \delta^{\star}\delta a)
\end{array}
\end{equation*}
\end{lemma}

\textit{Proof} is done by direct calculation, to illustrate we
show the check for $iii)$ which goes as follows
\begin{eqnarray*}
-\ih[p_{i}\psi^{i},a]_*  & = & -\ih (
p_{i}\psi^{i}a+(\frac{-i\hbar}{2}) (\frac{\partial^{2}}{\partial
y^{a}\partial \tilde{p}_{ a}} - \frac{\partial^{2}}{\partial
\tilde{y}^{a}\partial p_{ a}} )
p_{i}\psi^{i}a(x,\tilde{y},\tilde{p},\psi)\\ & & -
(-1)^{\tilde{a}} (a p_{i}\psi^{i}+(\frac{-i\hbar}{2})
(\frac{\partial^{2}}{\partial y^{a}\partial
\tilde{p}_{a}}-\frac{\partial^{2}}{\partial
\tilde{y}^{a}\partial p_{ a}}) a(x,y,p,\psi)\tilde{p}_{i}\psi^{i}) ) \\
&=&-\ih(\frac{-i\hbar}{2})(-\psi^{i}\frac{\partial a}{\partial
y^{i}}-(-1)^{\tilde{a}}\frac{\partial a}{\partial
y^{i}}\psi^{i})\\ &=& \psi^{i}\frac{\partial a}{\partial y^{i}}\\
&=&\delta a
\end{eqnarray*}
\qed

\bigskip

Note that property $iv)$ implies that for all $a\in\mc{W}$ there
is a decomposition
\begin{equation}
a=\delta \delta^{-1} a +\delta^{-1} \delta a+ a_{00}.
\label{deco1}
\end{equation}
The local expression of $\partial$ is
\begin{equation}
\partial a=\psi^{a} (\frac{\partial}{\partial x^{a}} + \Gamma_{ab}^{c}
p_{c}\frac{ \partial}{ \partial p_{b}} - \Gamma_{ab}^{c}
y^{b}\frac{ \partial}{ \partial y_{c}}) a. \label{rafai}
\end{equation}

\begin{lemma}It is possible to express $\partial a$ as
\begin{equation*}
\partial a=da+\ih [\Gamma,a]_*
\end{equation*}
for some  $\Gamma\in\mc{W}$ of odd parity.
\end{lemma}
\textit{Proof.} This can be shown as follows. Consider, for some
constant $\alpha$, the expression
\begin{equation*}
da+[\alpha\Gamma,a]_* = da+i\hbar\alpha(- \frac{
\partial\Gamma}{
\partial y^{b}}
 \frac{ \partial a}{ \partial p_{b}} +
 \frac{ \partial\Gamma}{ \partial p_{c}}
 \frac{ \partial a}{ \partial y^{c}}) + O(\hbar^{2}).
\end{equation*}
Comparing this equation with equation (\ref{rafai})  leads us to
the equations

\begin{equation*}
\psi^{a}\Gamma_{ab}^{c} p_{c}=-i\hbar \frac{ \partial
\alpha\Gamma}{
\partial y^{b}},\,\,\,\,\,\,\,\,
-\psi^{a}\Gamma_{ab}^{c} y^{b}=i\hbar\frac{
\partial \alpha\Gamma}{\partial p_{c}}.
\end{equation*}
This implies that we must take
$\Gamma=\Gamma_{ab}^{c}y^{b}p_{c}\psi^{a}$ and $\alpha=\ih$ and
the result follows. \qed
\bigskip

The auxiliary algebra is defined as the set
$\mc{W}_D:=\{a\in\mc{W}:D a=0\}$, for some connection $D=\psi^a
D_a$ that must satisfy the integrability condition $D^2=0$. It
turns out that taking $D=\partial$ is not a good choice as the
following proposition shows.
\begin{prop}
The integrability condition for the connection $\partial$ can be
expressed as
\begin{equation}
\partial^{2}a=\ih [R,a]_*
\end{equation}
where $R:=\half \psi^{b}\psi^{c} R_{abc}^{d}p_{d}y^{a}$.
\end{prop}
\textit{Proof.}
\begin{eqnarray*}
\frac{1}{2}[\partial,\partial]_* a  &=&\frac{1}{2}  [\psi^{a}
(\frac{\partial}{\partial x^{a}} + \Gamma_{ab}^{c} p_{c}\frac{
\partial}{ \partial p_{b}} - \Gamma_{ab}^{c} y^{b}\frac{
\partial}{ \partial y^{c}}),\psi^{d} (\frac{\partial}{\partial
x^{d}} + \Gamma_{de}^{f} p_{f}\frac{
\partial}{ \partial p_{e}} - \Gamma_{de}^{f} y^{e}\frac{
\partial}{ \partial y^{f}})]_*  a\\
&=&\frac{1}{2}\psi^a\psi^d ((\frac{\partial
\Gamma^f_{de}}{\partial x^a}-\frac{\partial
\Gamma^f_{ae}}{\partial x^d}) p_f \frac{\partial}{\partial p_e}
+(\frac{\partial \Gamma^f_{ae}}{\partial x^d}-\frac{\partial
\Gamma^f_{de}}{\partial x^a})y^e\frac{\partial}{\partial y^f} +\\
& & (\Gamma^f_{am}\Gamma^m_{de}-\Gamma^m_{ae}\Gamma^f_{dm}
)p_f\frac{\partial}{\partial p_e} +
(\Gamma^m_{ae}\Gamma^f_{dm}-\Gamma^m_{de}\Gamma^f_{am})y^e\frac{\partial
}{\partial y^f})a
\end{eqnarray*}
And since
\begin{equation*}
R^i_{jkl}=\frac{\partial \Gamma^i_{jl}}{\partial x^k}-
\frac{\partial \Gamma^i_{jk}}{\partial x^l} +
\Gamma^i_{mk}\Gamma^m_{jl}-\Gamma^i_{ml}\Gamma^m_{jk},
\end{equation*}
we have that
\begin{equation*}
\partial^{2}=\frac{1}{2}\psi^{b}\psi^{c}(R_{abc}^{d} p_{d}
\frac{\partial}{ \partial p_{a}}-R_{abc}^{d}y^{a}\frac{\partial}{
\partial y^{d}})
\end{equation*}
and on the other hand we have
\begin{equation*}
\ih[R,a]_* =\frac{\partial R}{\partial y^{a}}\frac{\partial a}{
\partial p_{a}}-\frac{\partial R}{
\partial p_{d}}\frac{\partial a}{
\partial y^{d}}.
\end{equation*}
This implies that we must take
\begin{equation*}
R=\frac{1}{2}\psi^{b}\psi^{c}R_{abc}^{d}p_{d}y^{a},
\end{equation*}
in order to have $\partial^{2}=\ih[R,\cdot\,]_* $. This completes
the proof. \qed
\bigskip

In other words the connection $\partial$ should be flat to fulfil
the condition, which is far too restrictive. The way out is to
define a new generalized connection of the form
\begin{equation*} Da=da+\ih
[-\psi^{a}p_{a}+\Gamma+\gamma,a]_* =
\partial a + \ih[-\psi^a p_a +\gamma,a]_*
\end{equation*}
where $\gamma$ is
\begin{equation}\label{gamma}
\gamma=\sum_{n=3}^{\infty}\Gamma^a_{i_1\ldots i_n,b} y^{i_1}\ldots
y^{i_n} p_a \psi^b,
\end{equation}
to be determined to fulfill the integrability condition. Further
calculation shows
\begin{eqnarray*}
D^{2}a &=& \ih[R,a]_* +\ih \partial [-\psi^{a}p_{a}+\gamma,a]_* +
\ih [-\psi^{a}p_{a}+\gamma,\partial a]_*  + (\ih)^{2}
[-\psi^{a}p_{a}+\gamma,[-\psi^{a}p_{a}+\gamma,a]_* ]_* \\ &=&
\ih[R,a]_*+\ih[\partial\gamma,a]_* +
(\ih)^{2}\frac{1}{2}[[\psi^{a}p_{a}+\gamma,-\psi^{a}p_{a}+\gamma]_* ,a]_* \\
&=& \ih[R+\partial\gamma-\delta\gamma+\ih\gamma\starl\gamma,a]_*
\end{eqnarray*}
Then the equivalent condition for having $D^{2}=0$ is
\begin{equation}\label{condi}
\delta \gamma = R+\partial\gamma+\ih \gamma^{2}
\end{equation}
\begin{prop}
$\gamma$ is a solution of equation (\ref{condi}) if and only if
\begin{equation}
\gamma=\delta^{-1}R+\delta^{-1}(\partial\gamma+\frac{i}{\hbar}
\gamma^{2}) \label{caracho}
\end{equation}
and the condition $\delta^{-1}\gamma=0$ is fulfilled.
\end{prop}
Fedosov's proof \cite{fedosov1} can be applied here so we shall
not include it.
\bigskip

Substitution of the general form of $\gamma$ given in
(\ref{gamma}) into equation (\ref{caracho}) leads to an iterative
process  with initial condition $\delta^{-1} R$. The first terms
of the solution are
\begin{equation*} \gamma=\frac{1}{3} R^d_{abc} y^a y^b \psi^c p_d
+ \frac{1}{12}
\partial_l R^d_{abc} y^l y^a y^b \psi^c p_d +\ldots.
\end{equation*}

The subalgebra $\mc{W}_D$ is defined by the condition $Da=0$,
i.e.,
\begin{equation}\label{nuevacondicion}
\delta a= \partial a+\ih [\gamma,a]_*
\end{equation}

\begin{prop}
There is a one to one correspondence
$\phi:\mc{O}_{\Omega^1M}[[\hbar]]\rightarrow \mc{W}_D$.
\end{prop}
\textit{Proof.} This can be shown as follows. Condition
 (\ref{nuevacondicion}) is equivalent to
\begin{equation}
a = a_{00} + \delta^{-1}(\partial a +[\frac{i}{\hbar}\gamma,a]_*
). \label{julio}
\end{equation}
The equivalence of these two equations is proved as in the Fedosov
construction. Indeed, since $D^{2}a=0$
\begin{equation}
\delta Da= \partial Da + [\ih\gamma,Da]_*  \label{talachas}.
\end{equation}
On the other hand using (\ref{julio}) $\delta^{ -1} Da =0$ and so
\begin{equation*}
Da=\delta^{ -1}(\partial Da+[\ih\gamma,Da]_* ).
\end{equation*}
Solution of this equation by an iterative process  implies that
$Da=0$. The converse assertion is trivial.

These are the first few terms of the solution for equation
(\ref{julio})
\begin{equation*}
a=a_{00}+\partial_i a_{00} y^i + \half \partial_i \partial_j
a_{00} y^i y^j +\frac{1}{6}\partial_i\partial_j\partial_k a_{00}
y^i y^j y^k -\frac{1}{12}R^{d}_{abc}y^a y^b p_d \frac{\partial
a_{00}}{\partial p_c}+\ldots
\end{equation*}
giving the one to one map $a_{00}\mapsto \phi(a_{00}):=a$. \qed

\bigskip
The star-product for $f_{00},g_{00}\in \mc{O}_{\Omega^1
M}[[\hbar]]$ is finally defined as
\begin{equation*}
f_{00}\starl' g_{00} =\phi^{-1}(f\starl g ),
\end{equation*}
where $\phi^{-1}(f\starl g ) =f\starl g (x,p,y,0)_{|y=0}$. The
star product $\starl'$ inherits from $\starl$ the natural shift
from left to right multiplication $f_{00}\starl
g_{00}=g_{00}\star_{-\hbar}f_{00}$, implying that our construction
will not depend on our choice to use left ideals instead of right
ideals.

\bigskip
We find our key result for this step
\begin{theo}\label{superlemma}
Let $f_{00},g_{00}\in \pi^{*}(\mc{O}_{M}[[\hbar]])$, then
\begin{equation*}
f_{00}\starl'g_{00}=f_{00} g_{00}
\end{equation*}
\end{theo}

\textit{Proof.} Suppose that in equation (\ref{julio}) the
starting condition $a_{00}(x,p)$ does not depend on $p$, therefore
the commutator $[\gamma,a_{00}]$ vanishes and one can check that
this happens for every step in the iterative solution. We are left
then with the equation
\begin{eqnarray*}
a&=&a_{00}(x,p)+\delta^{-1}(\psi^i \partial_i a).
\end{eqnarray*}
Solutions of this equation are
\begin{equation*}
a=a_{00}+\sum_{i=1}^{n}
\frac{1}{n!}\partial_{i_1}\ldots\partial_{i_n} a_{00}y^{i_1}\dots
y^{i_n}.
\end{equation*}
Star products of functions of this type are clearly just the usual
commutative product. Now let $f_{00},g_{00}$ be two functions not
depending on $p$ then
\begin{eqnarray*}
f_{00}(x)\starl'g_{00}(x)&=&\phi^{-1}(f\starl g)=\phi^{-1}(fg)\\
 &=& f_{00}(x)g_{00}(x).
\end{eqnarray*}
\qed

From this point we denote the star product on $\Omega^1 M$ simply
as $\starl$.
\begin{step}
\textbf{Define the left ideal $\mathbf{\mc{J}_{l}}$ and compute
the its normalizer $\mathbf{\mc{N}_{l}}$}.
\end{step}
With the natural projection $\pi:\Omega^1 M\rightarrow M$ the
ideal $I\subset \mc{O}_{M}[[\hbar]] $ can be lifted to $\Omega^1
M$ giving a set $\pi^{*}(I)\subset \mc{O}_{\Omega^1 M}$ which
defines a left ideal
\begin{equation*}
\mc{J}_{l}=\{\mc{O}_{\Omega^1 M}[[\hbar]]\starl \pi^{*}(I) \}.
\end{equation*}
Consider now the normalizer $\mc{N}_{l}\subset A_{\mc{M}}$ for the
left ideal $\mc{J}_{l}$,
\begin{equation*}
\mc{N}_{l}=\{ h\in \mc{O}_{\Omega^1 M}[[\hbar]]:\pi^{*}(I)\starl
h\subset\mc{J}_{l} \}.
\end{equation*}
Clearly $\mc{J}_{l}\subset\mc{N}_l $ and moreover $\mc{J}_{l}$ is
a double sided ideal of $\mc{N}_{l}$, this is
\begin{equation*}
f\starl s\in \mc{J}_l ,\,\,\,s\starl f\in\mc{J}_{l},
\end{equation*}
for all $f\in\mc{N}_{l},\, s\in\mc{J}_{l}$.

\begin{step}\textbf{Take the quotient}
$\mathbf{\mc{Q}_{X}:=\mc{N}_{l}/\mc{J}_{l}}$. The result is a well
defined non-commutative associative algebra which we call the
quantum algebra of observables of $X$.
\end{step}

Computing the normaliser of a one-sided ideal and taking the
quotient to find a non-commutative ring is a rather common
procedure which in general leads to conditions difficult to solve,
but in our case, remarkably we have
\begin{theo}
The algebra $\mc{Q}_{X}$ corresponds to the quotient solution
space of a finite number of partial differential equations and it
does not depend on the choice of generators of the ideal $I$.
\end{theo}
\textit{Proof.} The key point of the proof is given by theorem
(\ref{superlemma}). This implies that for any $g\in
\mc{O}_{\Omega^1 M}$ the condition to be in the normalizer
\begin{equation*}
\pi^{-1}(I)\starl g= 0 \,\, mod\,\,\mc{J}_l
\end{equation*}
is equivalent to
\begin{equation}
(\sum_{i=1}^{n}\alpha_i\phi_{i})\starl g =
\sum_{i=1}^{n}\alpha_i\starl (\phi_i\starl g)=
0\,\,mod\,\,\mc{J}_{l} \label{condition}.
\end{equation}
In other words to the condition that $\phi_i\starl
g=0\,\,mod\,\,\mc{J}_l$ for the $n$ generators $\phi_i$ of
$\pi^{-1}(I)$. Each equation being in fact a partial differential
equation for functions in $\mc{O}_{\Omega^1 M}[[\hbar]]$. The
independence of the choice of generators of the ideal follows
trivially from the fact that any new set of generators
$\tilde{\phi}_i$ can be rewritten as a combination of the original
ones leading again to equation (\ref{condition}). \qed

\section{Examples}
We shall develop next several examples of the explained technique
when the ambient configuration space is $\re^2 $. The natural
choice of star product is the Moyal product on $\Omega^1 \re^2$,
given by
\begin{eqnarray*}
f\moyall g &=& e^{ \lambda \sum_{i=1}^{2}( \frac{\partial
}{\partial_{x_{i}} } \frac{\partial }{\partial_{\tilde{p}_{i}} }-
\frac{\partial }{\partial_{\tilde{x}_{i}} } \frac{\partial
}{\partial_{p_{i}}})}
f(x_{1},x_{2},p_{1},p_{2})g(\tilde{x}_{1},\tilde{x}_{2},\tilde{p}_{1},\tilde{p}_{2})
_{|\tilde{x}=x,\tilde{p}=p}
\end{eqnarray*}
with $f,g\in \mc{O}_{\Omega^1 \re^2}[[\lambda]]$. Clearly the
moyal product of functions not depending on the momentum variables
coincides with the pointwise product. We shall be using the
following
\begin{lemma}
Every analytic function $f(x_1 ,x_2 ,p_1 ,p_2 )\in \Omega^1
\mathbb{R}^2$ can be uniquely decomposed as
\begin{equation}
f(x_1 ,x_2 ,p_1 ,p_2 )=\sum_{i,j=0}^{\infty} f_{ij} (p_1 , p_2
)\moyall x_1^i \moyall x_2^j
\end{equation} \label{decom}
\end{lemma}
\textit{Proof.} It is sufficient to note that
\begin{eqnarray*}
h(p_1 ,p_2 )x_{1}^{M}x_{2}^{N}&=& h(p_1,p_2)\moyall
 x_{1}^{M}x_{2}^{N}\\
& &-\sum_{n=1}^{M}\sum_{k=0}^{min(N,n)}\frac{(-\lambda)^{n}}
{(n-k)!k!}\frac{M!N!}{(M-n+k)!(N-k)!}x_{1}^{M-n+k}x_{2}^{N-k}
\frac{\partial^{n}h}{\partial p_{1}^{(n-k)}\partial p_{2}^{k}}
\end{eqnarray*}
implying that the Taylor expansion  can be re-expressed in terms
of the Moyal product. \qed

\subsection{The cross}
Consider now the analytic variety  of the cross defined by the
equation $x_1 x_2 = 0$. Any function on the quantum algebra
$\mc{Q}=\mc{N}_l / \mc{J}_l $  can be expressed as
\begin{equation*}
h(x_{1},x_{2},p_{1},p_{2})=h_{0}(p_{1},p_{2})+h_{1}(x_{1},p_{1},p_{2})
\moyall x_{1} +h_{2}(x_{2},p_{1},p_{2}) \moyall x_{2}.
\end{equation*}
The left ideal is
\begin{equation*}
\mc{J}_l =\{ f\moyall x_1  x_2 :f\in \mc{O}_{\Omega^1 \re^{2} }
\}.
\end{equation*}
The condition for the function $h$ to be in the normalizer
$\mc{N}_l $ is $x_1 x_2 \moyall h \in \mc{J}_l $, i.e.,

\begin{eqnarray*}
0 \; \mbox{mod}\mc{J}_l &=& x_1x_2\moyall h_0+x_1x_2\moyall h_1\moyall x_1+x_1x_2\moyall h_2\moyall x_2\\
 &=& [x_1x_2, h_0]+[x_1x_2, h_1]\moyall x_1+[x_1x_2, h_2]\moyall x_2\\
 &=& 2 \lambda \frac{\partial^2 h_0}{\partial p_1 \partial p_2} +
 ( \frac{\partial h_0}{\partial p_2}+\lambda \frac{\partial^2 h_1}{\partial p_1 \partial p_2}
 + x_1 \frac{\partial h_1 }{\partial p_2 } )\moyall x_1 \\
\rule{0pt}{4ex} && \qquad \qquad \qquad \qquad  +\,
(\frac{\partial h_0}{\partial p_1}+
 \lambda \frac{\partial^2 h_2}{\partial p_1 \partial p_2}
 + x_2 \frac{\partial h_2 }{\partial p_1 } )\moyall x_2
\end{eqnarray*}
Solutions have the form
\begin{eqnarray*}
h_1(x_1,p_1,p_2)&=&-\frac{h_0^2(p_2)}{x_1}+\xi(x_1,p_2)e^{-\frac{x_1p_1}{\lambda}}
+ a(x_1,p_1) \\
h_2(x_2,p_1,p_2)&=&-\frac{h_0^1(p_1)}{x_2}+\zeta(x_2,p_1)e^{-\frac{x_2p_2}{\lambda}}
+ b(x_2,p_2).
\end{eqnarray*}
where $h_0^1(p_1),h_0^2(p_2)$ are arbitrary functions. The
solutions of the form
$\frac{h_0^2(p_2)}{x_1},\frac{h_0^1(p_1)}{x_2}$ are not defined on
the cross, therefore we do not consider them, similarly the terms
$\xi(x_1,p_2)e^{-\frac{x_1p_1}{\lambda}}$ and
$\zeta(x_2,p_1)e^{-\frac{x_2p_2}{\lambda}}$ must be rejected as
they are not meromorphic in $\lambda$. (However such solutions may
have a physical interpretation which we hope to elucidate later.)

We have then a family of functions for the quantum algebra of the
cross given by
\begin{equation*}
\mc{Q}_C:=\{h\in \mc{O}_{\Omega^1
\re^2}[[\lambda]]:h(x_1,x_2,p_1,p_2)= a(x_1,p_1)\moyall x_1 +
b(x_2,p_2) \moyall x_2 \}
\end{equation*}
Computing the Moyal product of two elements $h,\tilde{h}\in
\mc{Q}_C $ and eliminating terms with the factor $x_1 x_2$ we find
\begin{eqnarray*}
h\moyall \tilde{h}&=&(a\moyall x_1 + b\moyall x_2)\moyall
(\tilde{a}\moyall x_1 + \tilde{b} \moyall x_2)\\ &=&a\moyall x_1
\moyall \tilde{a} + b \moyall x_2  \moyall \tilde{b}.
\end{eqnarray*}

In other words the quantum algebra of the cross has then elements
of the form
\begin{equation*}
h= (a(x_1,p_1), b(x_2,p_2))
\end{equation*}
where $a,b$ are arbitrary functions  and the noncommutative
product is
\begin{equation*}
h \star_C \tilde{h}= (a\moyall x_1 \moyall \tilde{a}, b \moyall
x_2 \moyall  \tilde{b}).
\end{equation*}

\subsection{The double line}

Consider now the analytic space of the double line defined by the
equation $x_2^2=0$. Functions of the quantum algebra $\mc{N}_l /
\mc{J}_l$ can be represented as $$ h(x_1 , x_2 , p_1 , p_2 )= h_0
(x_1 , p_1 , p_2 ) + h_1 (x_1 , p_1 ,p_2 )\moyall x_2. $$ The
condition for $h$ to be in the normalizer of the left ideal
$\mc{J}_l =\{f\moyall x_2^2 ; f\in \mc{O}_{\Omega^1 \re^2 } \}$ is
\begin{eqnarray*}
0\, mod \mc{J}_l
 &=& [x_2^2, h_0] + [x_2^2, h_1] \moyall x_2
\end{eqnarray*}
Leading to the differential equation
\begin{equation*}
x_2 \frac{\partial h_0 }{\partial p_2}+\lambda x_2
\frac{\partial^2 h}{\partial p_2^2}+\lambda^2 \frac{\partial^3
h}{\partial p_2^3}= 0
\end{equation*}
whose general solution is
\begin{equation*}
h(x_1 , x_2 , p_1 , p_2) = a(x_1 , p_1) + b( x_1 , p_1 ) p_2
+\left( c(x_1 , p_1 ) + d(x_1 , p_1)p_2 - \frac{b(x_1 , p_1)}{2
\lambda} p_2^2 \right) \moyall x_2.
\end{equation*}
The product of two of these functions can be represented as a
matrix star-product denoted as $\star$, in the following way
\[ \phi(h):= \left(
\begin{array}{cc}
a + 2\lambda d & b \\
2\lambda c &  a\
\end{array}
\right)\equiv \left(
\begin{array}{cc}
A & B \\
C &  D\
\end{array}
\right)
\]
Then
\begin{eqnarray*} \phi(h) \star \phi(\tilde{h})) &=& \left(
\begin{array}{cc}
A  & B \\
C & D\
\end{array}
\right) \moyall \left(
\begin{array}{cc}
\tilde{A} & \tilde{B} \\
\tilde{C} & \tilde{D} \
\end{array}
\right)\\
&=& \left(
\begin{array}{cc}
A\moyall\tilde{A}+B\moyall \tilde{C} &
A\moyall\tilde{B}+B\moyall\tilde{D} \\ C\moyall \tilde{A}+D\moyall
\tilde{C} & C\moyall \tilde{B}+D\moyall\tilde{D}
\end{array}
\right)
\end{eqnarray*}
is equal to $\phi(h\moyall \tilde{h})$.

\subsection{Line with a double point}

We shall proceed now with the quantization of the line with a
double point defined as the quotient of the quantum algebra of the
double line quotient by the ideal generated by $x_1 x_2=0$  which
has the form
\begin{eqnarray*}
\phi(h\moyall x_1 x_2) &=& \phi(h)\star \phi(x_1 x_2) \\
&=&\left(
\begin{array}{cc}
A & B \\
C &  D\
\end{array}
\right) \moyall \left(
\begin{array}{cc}
 0 & 0 \\
x_1 &  0\
\end{array}
\right) \\
&=& \left(
\begin{array}{cc}
 B\moyall x_1 & 0 \\
 D \moyall x_1 &  0\
\end{array}
\right),
\end{eqnarray*}
where $B,D$ are arbitrary functions depending on $x_1,p_1$. The
corresponding normalizer will be the set of solutions of the
equation
\begin{equation}\label{pasada}
\left(
\begin{array}{cc}
 0 & 0 \\
x_1 &  0\
\end{array}
\right) \moyall \left(
\begin{array}{cc}
a   & b \\
c &  d\
\end{array}
\right)=0\,\,\, mod\; \mc{J}_l,
\end{equation}
where $a,b,c,d$ are a different set of arbitrary functions
depending on $x_1,p_1$. This implies in turn that $b=0$ since the
only solutions of the differential equation $x_1 b+\lambda
\partial_{p_1}b=0$ are not in the space of acceptable formal
functions. Then  $a(x_1 , p_1)$ must be a solution of the equation
\begin{equation*}
x_1 \moyall a = 0\,\,  (mod\, D \moyall x_1).
\end{equation*}
Decomposing $a=\sum_{i=0}^{\infty} a_i (p_1 )\moyall x_1^i$ and
factoring out we can rewrite this as
\begin{equation*}
x_1 \moyall a_0 (p_1 ) = 0\,\,(mod\,D \moyall x_1).
\end{equation*}
\begin{lemma}\label{lemin}
\begin{equation*}
A(x,p)\moyall B(p)= \sum_{n=0}^{\infty} (2\lambda)^n A_n (p )
B^{(n)}(p)\,\, (mod\,D\moyall x)
\end{equation*}
where $A(x,p)=\sum A_n (p)\moyall x^n$ and $B^{(n)}$ is the $n$th
derivative of $B$.
\end{lemma}
\textit{Proof.}
\begin{eqnarray*}
\sum_{i=0}^\infty A_i (p)\moyall x^{i-1} \moyall x \moyall B(p)&=&
\sum_{i=0}^\infty A_i (p)\moyall x^{i-1} \moyall 2\lambda B^{(1)}\,\, (mod\,D\moyall x ) \\
&=&\sum_{i=0}^\infty (2\lambda)^i A_i (p)\moyall  B^{(i)} (p)
\end{eqnarray*}
\qed

In particular this implies that the second equation means that
$a=k$ for some constant. Then the  normalizer has the form
\begin{equation*}
\left(
\begin{array}{cc}
 k  &  0 \\
 c(x_1 ,p_1) &  d (x_1, p_1 )\
\end{array}
\right)
\end{equation*}
The last step is to factor out the members of the left ideal from
this normalizer. This means to take
\begin{equation*}
c(x_1 ,p_1 ) \,\,mod\;D\moyall x_1.
\end{equation*}

The resulting quantum algebra for the line with a double point is
the set of matrices of the form
\begin{equation*}
\left(
\begin{array}{cc}
 k   & 0 \\
 c(p_1) &   d(x_1 , p_1 ) \
\end{array}
\right)
\end{equation*}
with multiplication law
\begin{equation*}
\left(
\begin{array}{cc}
k   & 0 \\
c(p_1) &   d(x_1 , p_1 ) \
\end{array}\right) \star'
\left(
\begin{array}{cc}
\tilde{k}   & 0 \\
\tilde{c}(p_1) &   \tilde{d}(x_1 , p_1 ) \
\end{array}\right)
\end{equation*}
\begin{equation*}
 = \left(
\begin{array}{cc}
k \tilde{k}   & 0 \\
\tilde{k}c(p_1)+\sum_{n=0}^{\infty} (2\lambda)^n d_n(p_1
)\tilde{c}^{(n)}(p_1)& d(x_1 , p_1 )\moyall \tilde{d}(x_1 , p_1 )
\
\end{array}\right)
\end{equation*}
where we have used lemma \ref{lemin}.


\subsection{The doubly fattened circle}

Let us consider now the space associated to the ideal generated by
$x_1^2 + x_2^2$, that we refer to as the "doubly fattened" circle.
Although the zero set of this  polynomial is just the origin the
resulting quantum algebra, as we shall show, is nontrivial. Any
function in this quantum algebra can be represented as
\begin{equation*}
h(x_1,x_2,p_1,p_2)=h_0(x_1,p_1,p_2)+h_1(x_1,p_1,p_2)\moyall x_2
\end{equation*}
The left ideal is
\begin{equation*}
\mc{J}_l=\{f\moyall (x_1^2+x_2^2): f\in \mc{O}_{\Omega^1
\re^2}[[\lambda]]\}.
\end{equation*}
The condition for a function $h(x_1,x_2,p_1,p_2)$ to be in the
normalizer is
\begin{eqnarray*}
0\;mod\,\,\mc{J}_l&=& (x_1^2+x_2^2)\moyall h\\
&=& -\frac{\partial h_1}{\partial p_2 }x_1^2 +2\lambda
x_1\frac{\partial^2 h_1}{\partial p_1
\partial p_2} - \lambda^2 \frac{\partial^3 h_1}{\partial p_1^2
\partial p_2}+ \lambda\frac{\partial^2 h_0}{\partial
p_2^2}+x_1\frac{\partial h_0}{\partial p_1}\\ & & +(\frac{\partial
h_0}{\partial p_2} + \lambda \frac{\partial^2 h_1}{\partial p_2^2}
+ x_1\frac{\partial h_1}{\partial p_1})\moyall x_2
\end{eqnarray*}
Manipulations in the last equation lead to the condition
\begin{equation*}
x_1^2(\Delta h_1)-\lambda^2 \frac{\partial^2}{\partial
p_2^2}(\Delta h_1)=0,
\end{equation*}
where $\Delta=\partial^2_{p_1}+\partial^2_{p_2}$, is the Laplacian
operator. Therefore the two dimensional spherical harmonics give a
family of solutions $h_1$.

\subsection{Conclusion and Acknowledgments}

 We have shown a way to define non-commutative associative
 products to analytic spaces immersed in analytic manifolds.
 The procedure works for smooth spaces and more remarkably
 it works for singular spaces. This opens  new possibilities
 in the field of deformation quantization and leaves many open  questions.
 One such question is how the singularity affects the resulting algebra,
 or more generally how the singularity type affects it.
 Questions like these can be studied through the introduction of parameters
 deforming the analytic spaces which will enter the partial differential equations
 defining the associated algebras. These equations turn
 out to be complex in most cases. New techniques need to
 be developed to find the representation of the algebras to make possible
 the study of such questions.

 Another interesting possibility is to develop a comparative study of
 the different quantization programmes.
 The common goal of all such programmes is to give a quantized
 version of classical physical systems. Naturally for any given
 classical system there should be a unique physical quantum version, thus
 one would expect the different approaches to be equivalent in some sense.
 Having an affirmative answer to this would give a hint of
 some deep mathematical relations between the different approaches.

 Another important line of work is to develop proper physical
 applications of the programme. A rather interesting question to be studied is
 to find out how the singularity affects the physics of the space.

 We expect to elucidate these an other questions in the future.

\bigskip
 The author would like to give special thanks Prof. Sergei Merkulov for all his
 valuable advices. Support from CONACYT-Mexico (grant 120-841) is
 gratefully acknowledged.

\end{document}